# Oxidation of graphite surface: the role of water

D.W. Boukhvalov[1,2]

[1]*Department of Chemistry, Hanyang University, 17 Haengdang-dong, Seongdong-gu, Seoul 133-791, Korea*

[2]*Theoretical Physics and Applied Mathematics Department, Ural Federal University, Mira Street 19, 620002 Ekaterinburg, Russia*

*Based on density functional calculations, we demonstrate a significant difference in oxidation patterns between graphene and graphite and the formation of defects after oxidation. Step-by-step modeling demonstrates that oxidation of 80% of the graphite surface is favorable. Oxidation above half of the graphite surface significantly decreases the energy costs of vacancy formation with $CO_2$ production. The presence of water is crucial in the transformation of epoxy groups to hydroxyl, the intercalation with further bundle and exfoliation. In water-rich conditions, water intercalates graphite at the initial stages of oxidation and oxidation, which is similar to the oxidation process of free-standing graphene; in contrast, in water-free conditions, large molecules intercalate graphite only after oxidation occurs on more than half of the surface.*

E-mail: danil@hanyang.ac.kr

## 1. Introduction

Graphite oxide (GO) was first synthesized more than 150 years ago[1] and became an interesting subject in current chemistry and physics in 2006[2] after the graphene breakthrough.[3] At present, this compound was proposed to be a source of graphene by exfoliation[4] and reduction[5] and as a material for catalyst,[6,7] filters,[8,9] solar cells,[10] magnetic materials,[11] ink-jet printing of electronic devices[12] and various biological applications.[13] These great prospects of GO require improving its large-scale fabrication methods.

There are two primary feasible methods for the large-scale GO production: Brodie[1] and Hummers.[14] Other methods, such as epitaxial graphene oxidation,[15] appear unsuitable for large-scale production because of the method difficulties and low yield. Although hundreds of scientific papers examined GO synthesis, details of the oxidation process remain unclear. Based on the sum of experimental papers, we conclude that compared to GO synthesized using the Brodie method, GO synthesized using the Hummers method has many more epoxy groups than hydroxyl groups, higher levels of carbon oxidation[16-18] and greater numbers of visible holes in the graphene sheets.[19] There is also other functional groups on GO reported experimentally but the content of these species is rather low and in the models usually considered only two most propagated groups.

Recent theoretical works discuss the effects of the atomic structure on the electronic,[16,20,21] optical,[22] mechanical,[23] properties of GO and further decoration of oxidized graphene,[24] the re-oxidation of the reduced GO,[7] the role of water in the stability of several configurations and stacking order[25] and the formation of capillaries in the graphene monolayer oxidation process.[9] Recent experimental[26,27] and theoretical works[27] demonstrate that the oxidation of graphene bi- and trilayers is different from that of monolayer. These results demand the modeling of graphite oxidation to build a proper model of the GO formation process. Our modeling of the hydrogenation of polycyclic aromatic hydrocarbons[28] and the hydrogenation and oxidation of graphene sheets[27,29] demonstrate that only the step-by-step modeling correctly explains the experimental results. Recent experimental works demonstrate the main steps of graphite oxidation: first intercalation with further oxidation.[30] But intercalations of graphite usually require high temperatures[31] in contrast to the room temperature formation of GO.

Another unclear on atomic level phenomenon of graphite oxidation is the formation of cracks and holes in graphene sheets with CO and $CO_2$ production. In the previously proposed model, the unzipping by line formation along zigzag directions[33] is only achieved in nanotubes[33] nanographenes;[34] in large graphene sheets, the graphite formation of these lines is energetically unfavorable.[29] In our recent work,[27] we proposed a new model of hole formation by removing the carbon atom connected with the epoxy group, which formed a monovacancy and a carbon dioxide molecule (Fig. 1a).

In this work, we will study the energetics of the step-by-step oxidation process of graphene mono- and six-layer. To imitate the oxidation of graphite by mineral acids, we study the one-by-one addition of epoxy groups on one side of graphene and graphite (Figs. 1a, c and e). We also explore the activation with $CO_2$ formation (Figs. 1b and f), the transformation of epoxy groups to hydroxyl in the presence of water (Figs. 1c and d) and the energetics of water and sulfur acid (as an example of large molecule in the GO fabrication process) intercalation under the top graphite layer (Fig. 2) as a function of the graphite surface. Presented modeling is rather idealized and does not provide explanation of whole graphite oxidation process but discuss several issues of graphene and graphite stability in oxidative environment that can be helpful for improvement of the quality of anti-corrosion carbon coverages, graphite-based anodes of batteries and perhaps improvement of the methods of GO production.

## 2. Computational method and model

We used the pseudo-potential code SIESTA[35] to perform the density functional theory (DFT) calculations (refer to other works[7,9,20,27,29,36] for details). All calculations were performed using the local density approximation (LDA),[37] which is feasible to model the graphite surface functionalization.[27,36] Other alternative approach for the modeling of weakly bonded layered systems is GGA+vdW.[38-40] There is several different approaches that include van der Waals forces but all of them lack for description of graphite some underestimate[38] other overestimate[39] binding energies and interlayer distances in graphite. The varying of interlayer distance is in range of 0.2 Å that is in order of out-of-plane lattice distortions of functionalized graphene[41] and can be crucial for the modeling of graphite oxidation and the values of interlayer binding energies vary with magnitude of 10 meV per carbon-carbon bond that is also in order of binding energies of water intercalated oxidized graphite (se discussion below). For additional check we performed the calculation of these values by GGA+vdW approach realized in used code[40] and find interlayer distances 3.52 Å and binding energy 28 meV per bond instead 3.33 Å[43] and 37 meV[43] in experiment and 3.36 Å and 34 meV calculated by LDA.

To model the graphite surface, we used a supercell of six-layered graphene (see discussion in Ref. [36]) with 32 carbon atoms in each layer. The atomic positions were fully optimized. The wave functions were expanded with a double-$\zeta$ plus polarization basis of localized orbitals for carbon and oxygen and double-$\zeta$ for hydrogen. The force and total energy was optimized with an accuracy of 0.04 eV/Å and 1 meV, respectively. All calculations were performed with an energy mesh cut-off of 360 Ry and a **k**-point mesh of 8×8×4 in the Monkhorst-Pack scheme.[44] The formation energies for the oxidation were calculated using a standard equation: $E_{chem} = (E_{host+guest} - [E_{host} + E_{guest}])$, where $E_{host}$ is the total energy of the system before adsorption or intercalation, and $E_{guest}$ is the total energy of molecular oxygen in the triplet state divided by two or the energy of water of sulfur acid molecules in an empty box.

## 3. Results and discussion

### 3.1. Oxidation of graphene and graphite surfaces

First, we compared the energetics of graphene monolayer and graphite oxidation (Figs. 3a and b). There are two primary differences. First, the first steps of the graphite oxidation require considerably higher energy than those of graphene. For the check of possible effect of the number of layers usen in our mode we performed the calculations of formation energy of first step of oxidation also for ten layers of graphene in slab and obtain the same value as for six layers. These results are qualitatively consistent with experimental results, which demonstrate the oxidation of graphene and graphene trilayer from air at 200°C and 700°C, respectively.[26] Second, graphene favors 100% oxidation on the surface (Fig. 1e), whereas graphite favors approximately 60% oxidation on the surface. Both differences make the free-standing graphene membrane more flexible than graphite when the top layer binds with the bottom one via π-π bonds. The transformation of epoxy groups to hydroxyl in the presence of water (Figs. 1c and d) below 40% oxidation level is energetically favorable for graphene and graphite. Further formation of hydroxyl groups on the graphene monolayer is energetically favorable because of significant lattice distortions of the –OH groups,[20] whereas for the graphite surface, this process continues until approximately 80% oxidation level because the interaction of the top graphite layer with the substrate decrease the lattice distortions. This results is in qualitatively agreement with experimental observation of larger number of hydroxyl groups in GO samples fabricated by Brodie method when water added after oxidation.[45]

### 3. 2. Activation of graphene and graphite surfaces

The activation of graphene and graphite (Figs. 1b and f) also depends on the oxidation level. For the graphene monolayer, the activation energy significantly decreases at the oxidation level of 20% and becomes almost zero at the oxidation level of 40 to 60%. For graphite, surface activation is only possible at the oxidation levels of approximately 50% and above 80% when this process becomes exothermic because dangling bonds that appear after activation form the bonds with the carbon atoms from the bottom layers (see Fig. 1f). These results are qualitatively consistent with the experimental results,[26,27] which demonstrate that the graphene monolayer has a higher level of perforation than the graphene trilayer. Thus, from our modeling results, we conclude that for graphite, surface oxidation without vacancy formation is possible up to approximately 50%, and further oxidation, activation and transformation of epoxy groups to hydroxyl in the presence of water are possible at all oxidation levels below 80%.

**3.3. Intercalation of graphite**

Our previous work demonstrates that the interlayer bonds in graphite weaken with the increase in surface functionalization level.[36] Based on this result, the next step of our survey is to determine the relationships between the surface oxidation and the required energetics of intercalation of water or larger molecules to delaminate or peel off the oxidized top layer and oxidize the next layer of graphite. For additional check we perform the calculations of binding energies in several configurations with using GGA+vdW approach[40] and find that the values of binding energies decrease in order 7-12 meV/bond. For the check of the effect of intercalates concentration we double the number of water molecules and find decay of binding energy at 5 meV/bond. Our calculation results (Fug. 3c) demonstrate that the intercalation by sulfuric acid (Fig. 2c) is only possible after the top graphite layer is 60% oxidized. Unlike the large $H_2SO_4$ molecules, intercalation by water (Fig. 2a) can be realized in the first oxidation steps. The transformation of epoxy groups to hydroxyl forms hydrogen bonds between these groups and other water molecules, which additionally facilitates the intercalation process. This result is qualitatively consistent with the significant increase of the amount of water between the layers of the graphene oxide bilayer when the top layer interacts with water.[19]

Based on these results we can speculate about relations between oxidation and intercalation of graphite at room temperature in GO formation process.[30] Intercalation unoxidized graphite requires rather high energies even for such small molecules as water and for larger molecules such as $H_2SO_4$ require energies about four times bigger (Fig. 3c). Perhaps the oxidation of narrow areas in vicinity of graphite boundaries facilitates penetration of first molecules between graphene edges. This near boundaries intercalation should provide increasing of interlayer distances in graphite that decrease binding energy between layers and makes possible room temperature intercalation prior oxidation. After intercalation shifted up graphene layers became achievable for oxidative species from both sides and the oxidation of graphene sheet continues. This process could forms capillary-like patterns along a zigzag direction, which enables water to flow in the interlayer space of GO.[9]

In absence of preliminary intercalation and water, oxidation occurs in the step-by-step functionalization of the graphite surface because of the epoxy groups. When more than 50% of the surface is oxidized, the vacancy formation process starts, which can be interpreted as the starting point of large-hole formations and cracking of the graphene sheets. Further oxidation of non-intercalated graphene in water-free conditions decreases the required energy for intercalation by large molecules and peeling of the top oxidized and perforated graphene layer. Thus, without preliminary intercalation and in water-free conditions, should provide layer by layer peeling of one-side oxidized graphene sheets from graphite that is in disagreement with experimentally observed oxidation without exfoliation. Well then the modeling of graphene oxide fabrication must discuss oxidation of intercalated graphite instead layer-by-layer oxidation of graphite or graphene monolayer.

## 4. Conclusions

In summary, based on the DFT modeling results, we demonstrate the significant qualitative differences between graphene monolayer oxidation and graphite surface oxidation. This result demonstrates that the model of free standing graphene oxidation is limitedly valid for description of graphite oxide formation. For graphite, the presence of water is crucial in the oxidation and exfoliation process. With water, the epoxy groups transform into hydroxyl and water intercalates under top layer of graphite at the initial oxidation steps. Further oxidation of the shifted up layer can be described as the oxidation of graphene monolayer. Without water, a significant oxidation level of the graphite surface (up to 80%) can be achieved. When the graphite surface is more than 60% oxidized, the vacancy formation and intercalation by larger molecules such as sulfuric acid begin. The atomic structure of graphene oxide formed without water should be different. The performed modeling of graphite oxidation and exfoliation in various conditions can be used to improve further modeling graphene oxide production process and demonstrate the need of taking into account all details.

**Acknowledgements** The work is supported by the Ministry of Education and Science of the Russian Federation, Project N 16.1751.2014/K

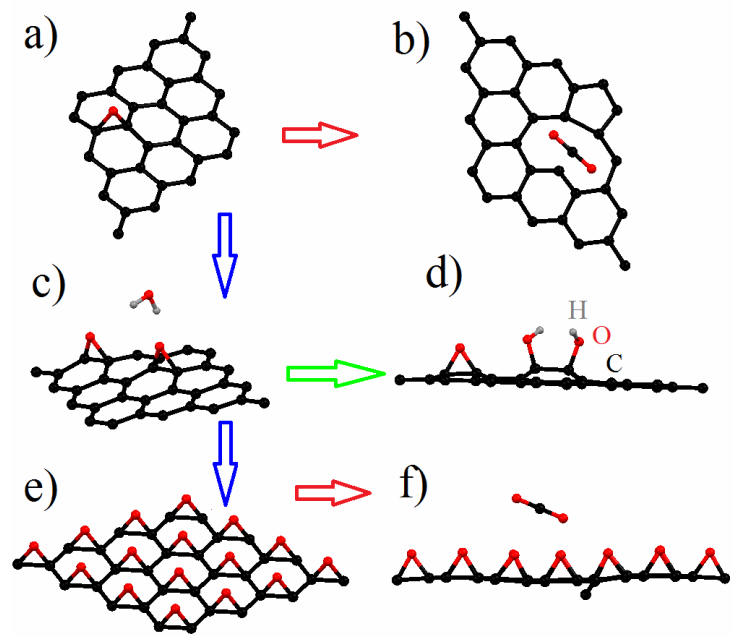

**Figure 1** Optimized atomic structures of the top graphite layer after the adsorption of the first oxygen atom (a) and the formation of monovacancy (activation) and $CO_2$ molecule (b), adsorption of the second oxygen atom (c) and transformation of epoxy groups to hydroxyl in the presence of water molecules (d); totally oxidized top layer of graphite before (e) and after (f) activation.

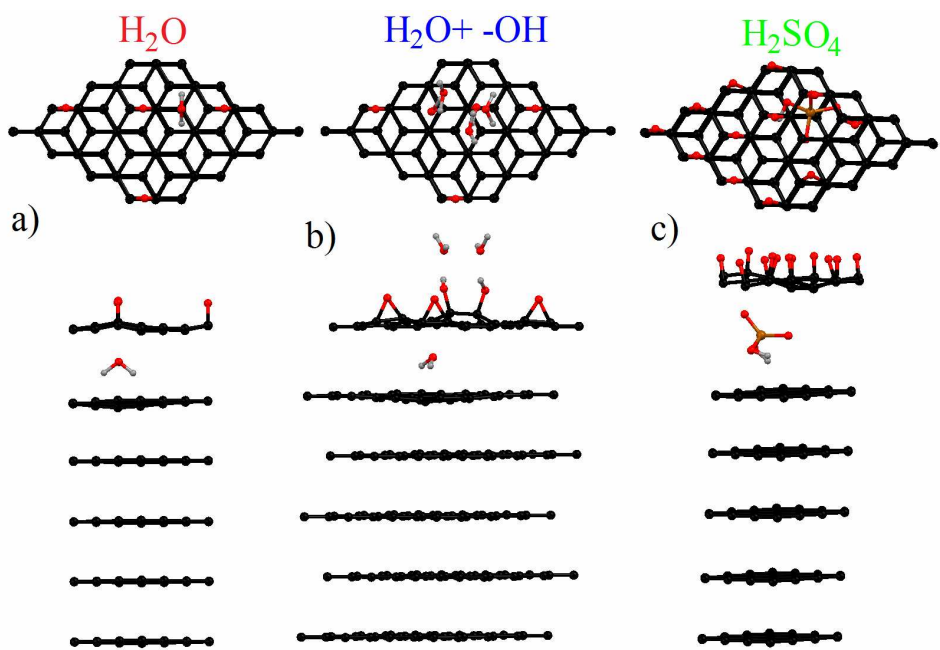

**Figure 2** Top and side views of graphite with 25% oxidation of the top layer and intercalated water molecule (a), graphite with the same oxidation level after the epoxy groups transform into hydroxyl (b) and graphite with 75% oxidation of the top layer and intercalated $H_2SO_4$ molecule (c).

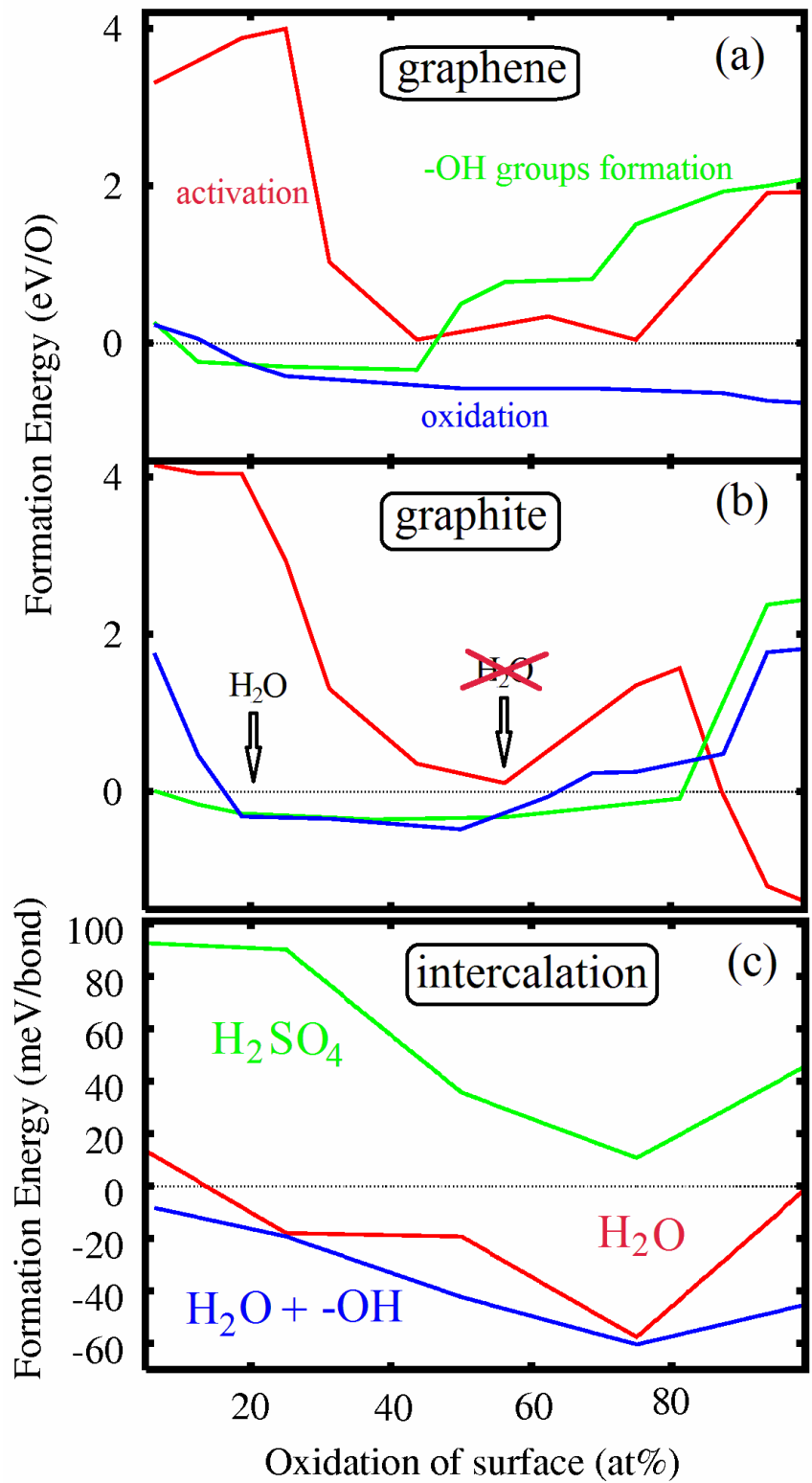

**Figure 3** Formation energies to oxidize (Figs. 1a and e), activate (Figs. 1b and f) and transform epoxy groups to hydroxyl in the presence of water (Figs. 1c and d) for graphene mono (a) and six-layer (b). The formation energy per carbon-carbon interlayer bond for three different intercalation scenarios is shown in Fig. 2. The arrows on panel (b) indicate the probable beginning of intercalation in water-rich and water-free conditions.